\documentclass[11pt,a4paper,oneside,openany]{article} 

\usepackage[top=2.3cm,left=2.3cm,right=2.3cm,bottom=2.3cm]{geometry}
\usepackage{graphicx}
\usepackage{mathtools}
\usepackage{physics}
\usepackage{amssymb}
\usepackage{nicematrix}
\usepackage{float}
\usepackage{empheq}
\usepackage[most]{tcolorbox}
\usepackage[utf8]{inputenc}
\usepackage{pifont}
\usepackage{subcaption}

\usepackage{tabularx}

\usepackage{authblk}   
\usepackage{hyperref}  

\newtcbox{\mymath}[1][]{%
	nobeforeafter, math upper, tcbox raise base,
	enhanced, colframe=blue!30!black,
	colback=blue!30, boxrule=1pt,
	#1}
\newcommand{\e}[1]{\mathrm{e}^{#1}}

\newcommand{\sgn}[1]{\mathrm{sgn}({#1})}

\makeatletter
\renewcommand\AB@affilsepx{\protect\\}

\makeatother

\title{
Symmetry‑Guided Design of Quantum Couplers\\ in Dirac materials: AA‑Bilayer Graphene Coupler}

\author[1,2]{Petr Červenka\thanks{\href{mailto:cervep12@fjfi.cvut.cz}{cervep12@fjfi.cvut.cz}}}
\author[1]{Vít Jakubský\thanks{\href{mailto:jakubsky@ujf.cas.cz}{jakubsky@ujf.cas.cz}}}

\affil[1]{\it Nuclear Physics Institute, Czech Academy of Sciences\\
250 68 \v{R}e\v{z}, Czech Republic}

\affil[2]{\it Faculty of Nuclear Sciences and Physical Engineering\\
Czech Technical University in Prague}

\begin{document}

\maketitle

\begin{abstract}
We develop a theoretical framework for designing quantum couplers based on Dirac materials that can modulate the polarization of transmitted quasiparticles without significantly perturbing their propagation. We analyze in detail the conditions required for perfect transmission (Klein tunneling) together with controlled polarization transformation of the incoming states.
We then discuss an explicit model of a quantum coupler composed of AA‑stacked bilayer graphene nanoribbons with armchair edges and a localized interlayer interaction. Perfect transmission through the desired polarization channels is examined for both narrow and wide couplers. We show that the transmission of polarized states can be finely tuned by external fields.

\end{abstract}
\section{Introduction}


For practical applications, it is essential to transport quantum states with a well‑defined polarization and, when required, to modify that polarization in a controlled and predictable manner. Quantum couplers can serve as key electronic components that enable these operations.

Two‑dimensional Dirac materials offer multiple mechanisms for polarizing quasiparticles, each associated with specific interactions capable of modifying that polarization. In graphene, Dirac fermions can be polarized with respect to their spin degree of freedom. Spintronics exploits this property in electronic applications, employing interactions such as spin–orbit coupling as a means of control \cite{Pesin2012, Han, Kochan, Hu}.

The presence of two inequivalent Dirac points in graphene’s band structure gives rise to valleytronics \cite{Tworzydlo, Schaibley, Vitale2018}. This field relies on the fact that valley polarization is preserved in the absence of intervalley‑mixing interactions, such as Kekulé distortions \cite{GalvanGarcia2022} or strain‑induced effects \cite{Altland2006}. The valley degree of freedom also plays an important role in Andreev reflection at graphene–superconductor interfaces, where particle–hole polarization becomes relevant \cite{Beenaker_2006, Andreev_reflection, Andreev_reflection2}, for example in quasiparticle lensing phenomena.

Parallel layers of two‑dimensional materials can interact through van der Waals–type forces \cite{Geim}. The presence of additional layers introduces new possibilities for quantum transport and enables new forms of quasiparticle polarization \cite{Abdullah1, Abdullah2, blisters, Abdullah3, sanz, Mirzakhani, Jaskolski2018}.

Transmission through van der Waals domain walls arising from fluctuations of the interlayer interaction in AA‑ and AB‑stacked bilayer graphene was analyzed in detail in \cite{Abdullah1}. Band‑gap opening in AA‑stacked bilayers induced by dielectric environments, and the resulting transport properties of quasiparticles across such domains, were studied in \cite{Abdullah2}. It was shown in \cite{blisters} that blisters in bilayer graphene can confine charge carriers, while bilayer interfaces can act as collimators for wave packets \cite{Abdullah3}. Crossings of armchair nanoribbons with AA‑type interlayer coupling at the junction were proposed as beam splitters and mirrors for electron optics \cite{sanz}.
Layer decoupling at the edges of bilayer nanoribbons can generate valley‑polarized domain‑wall states \cite{Mirzakhani}. Depending on the strength and spatial profile of the interlayer interaction, quasiparticles may become localized predominantly on a single layer, giving rise to layer engineering—or layertronics—as a functional degree of freedom \cite{Jaskolski2018}. 

 Klein tunneling is a phenomenon in which the Dirac fermions transmit through potential barriers without being reflected \cite{Katsnelson_Klein}. It compromises confinement of quasi-particles by electric field in graphene and therefore, makes its direct use in electronic devices rather indirect. Klein tunneling in bilayer graphene was analyzed e.g. in \cite{Abdullah1}, \cite{Sanderson},  where perfect transmission through the layer-polarized or cone-index polarized channels was discussed. In \cite{Abdullah2}, Klein tunneling was considered through the barrier formed by interaction with dielectric medium that opens the gap in the spectrum.

In this article, our central objective is to develop theoretical guidelines for designing quantum couplers capable of switching or modulating polarization without significantly disturbing the transmitted particles. The key ingredient for achieving a low‑loss device is the exploitation of Klein tunneling \cite{BetancurOcampo2025KleinReview}. In graphene, this effect limits electrostatic confinement; however, it can be turned into an advantage in devices where reflection suppression is desirable. By operating under conditions in which Klein tunneling minimizes backscattering, a coupler can transmit wave packets with negligible reflection while performing the required polarization transformation.

We focus specifically on the theoretical design of a multilayer graphene coupler composed of parallel graphene nanoribbons with local AA‑bilayer stacking and armchair edges. Nevertheless, many of the underlying principles naturally extend to other geometries and material platforms.

\section{Polarization and Klein tunneling in couplers\label{section2}} 
Let us consider a system described by the following Hamiltonian 
\begin{equation}
H=-i\gamma_x\partial_x-i\gamma_y\partial_y+V_0(x)+V_{cpl}(x)=H_{out}+V_{cpl}.\label{initH}
\end{equation}
Here, $H_{out}$ describes dynamics outside the coupler, whereas $V_{cpl}$ corresponds to the interaction represented by the coupler. We assume that $V_{cpl}$ is vanishing outside the interval $x\in[0,L]$. The Hamiltonian commutes with $p_y=-i\partial_y$, $[H,p_y]=0$. We leave the width of the coupler and related boundary conditions unspecified at the moment.

We assume that the system outside the coupler has a symmetry $S$, $[H_{out},S]=0$ that can be used to fix polarization of the quantum states outside the coupler. 
The common eigenvectors of $H_{out}$, $p_y$, and $S$,
\begin{equation}
H_{out}\psi_{k\mu}=E\psi_{k\mu},\quad p_y\psi_{k\mu}=k\psi_{k\mu},\quad S\psi_{k\mu}=\mu\psi_{k\mu}.
\end{equation}
shall form the basis of the Hilbert space. We say that a state $\psi$ is polarized with respect to $S$ provided that it is an eigenvector of the operator. 

Let us suppose that a plane wave with given polarization $\psi_{in}=\psi_{k\mu}$ bounces on the coupler from the left. The transmitted wave $\psi_t$ can be written as a linear combination of polarized states $\psi_t=\sum_{\nu}c_{\nu}\psi_{k\nu}$. The explicit form of the polarization parameters $c_{\nu}$ are determined by the actual form of $V_{cpl}$ and computing them is essentially equivalent to solving the full scattering problem. As a result, analyzing how the coupler modifies polarization is, in general, a nontrivial task.

We want to gain insight into the operation of the coupler  without the need to specify $V_{cpl}$ explicitly. To this aim, we propose constructing an auxiliary operator $\tilde{H}$ that approximates the full Hamiltonian $H$ in the low‑energy regime relevant for quantum transport, while admitting eigenstates that can be analyzed in a straightforward manner.  

Let us consider a unitary operator $U$ that satisfies
\begin{equation}
\quad [U(x),\gamma_x]=0,\quad [U(L),V_0(x)]=0, \quad U=\begin{cases}\mathbb{I},\quad x<0,\\
U(x ),\quad x\in[0,L],\\
U(L),\quad x>L.\end{cases}\label{Utrans}
\end{equation}
We define the auxiliary Hamiltonian $\tilde{H}$ as follows,
\begin{eqnarray}\label{Htilde}
\tilde{H}\equiv UH_{out}U^{\dagger}&=&-i\gamma_x\partial_x-i\gamma_y\partial_y-i\gamma_xU(\partial_xU^\dagger) -i[U,\gamma_y]U^\dagger \partial_y+UV_0U^\dagger\nonumber
\\
&=&H_{out}+V_{cpl}-i[U,\gamma_y]U^\dagger \partial_y.
\end{eqnarray}
The potential $V_{cpl}$ reads as
\begin{equation}\label{Vcpl}
V_{cpl}\equiv -i\gamma_xU(\partial_xU^\dagger)-V_0+UV_0U^\dagger.
\end{equation}
It is hermitian and vanishing outside the interval $[0,L]$.
The eigenstates $\tilde{\psi}$ of $\tilde{H}$ can be found directly as $\tilde{\psi}=U\psi$ from  eigenstates $\psi$ of $H$. These states are continuous at $x=0,L$ by construction\footnote{When assuming continuity of $\psi$.}.

The operators $\tilde{H}$ and $H$ coincide for $x<0$, but can differ elsewhere. The deviation is represented by the last term in (\ref{Htilde}) that depends on the transversal momentum $p_y=-i\partial_y$. The operator $\tilde{H}$ can provide a good approximation of $H$ in the subspaces where the transversal momentum acquires small values. A simple but important example is the situation where the states are incoming with normal incidence to the coupler,  
i.e., they satisfy $p_y\psi=0$. Then the correction term vanishes and $\tilde{H}$ describes the exact dynamics of the particles passing through the coupler. In this subspace, there holds
\begin{equation}
\tilde{H}_0\equiv \tilde{H}_{p_y=0}= H_{p_y=0}=-i\gamma_x\partial_x+V_0+V_{cpl},
\label{tildeH0}
\end{equation}
where the interaction $V_{cpl}$ corresponding to the coupler is then given by \eqref{Vcpl}. Eigenvectors of the operator can be obtained directly as \begin{equation}\tilde{H}_0\tilde{\psi}=E\tilde{\psi},\quad \tilde{\psi}=U\psi=\begin{cases}\psi(x), & x<0\\U(x)\psi(x),& x\in[0,L]\\U(L)\psi(x),& x>L. \end{cases} \end{equation}
In this regime, it is straightforward to find the polarization coefficients $c_\nu$ of the transmitted wave. They are encoded in the matrix $U(L)$,
\begin{equation}
\psi_t(x)=U(L)\psi_{in}(x)=\sum_{\nu}c_\nu\psi_{k\nu}.
\end{equation}
This relation can be used to reverse‑engineer the coupler: it allows us to choose the matrix $U$
such that the transmitted wave acquires the desired polarization. Once $U$
is fixed, it determines the corresponding potential $V_{cpl}$ that performs the required operation. This approach is exact within the subspace $p_y=0$. In the next section, we show that this subspace can play a major role in transport properties under realistic conditions.

Let us fix $H_{out}$ as the energy operator of the free Dirac particle in graphene,
\begin{equation}H_{out}=-i\sigma_0\otimes\sigma_1\partial_x-i\sigma_0\otimes\sigma_2\partial_y.\end{equation}
In the tensor product, the right matrix acts on pseudo-spin whereas the left matrix can act e.g. on spin-, valley- or layer-degree of freedom. In the context of the current work, we will consider the last one. 

The most general matrix $U$ that satisfies the commutator \eqref{Utrans} is 
 \begin{equation}   U(x)=\sum_{\substack{\mu=0}}^4\left(\alpha_{\mu 0}(x)\sigma_{\mu}\otimes\sigma_0+\alpha_{\mu 1}(x)\sigma_{\mu}\otimes\sigma_1\right)\label{Upolarization}\end{equation}
where the coefficients $\alpha_{\mu j}(x)$ are such that the matrix is unitary and $U(x<0) = \sigma_0\otimes\sigma_0$. The coupler is represented by the potential $V_{cpl}=-i\gamma_xU(\partial_xU^\dagger)$ whose matrix structure can be found in the following form
\begin{equation}
V_{cpl}(x) = 
\begin{cases}
v_{00}(x)\sigma_0\otimes\sigma_0
+ v_{30}(x)\sigma_3\otimes\sigma_0 +
    &\qquad \text{(I) electric field} \\[6pt]

+v_{01}(x)\sigma_0\otimes\sigma_1
+ v_{31}(x)\sigma_3\otimes\sigma_1+
    &\qquad \text{(II) magnetic field} \\[6pt]
+v_{10}(x)\sigma_1\otimes\sigma_0
+ v_{20}(x)\sigma_2\otimes\sigma_0+
    &\qquad \text{(III) AA-type  interaction } \\[6pt]
+v_{11}(x)\sigma_1\otimes\sigma_1
+ v_{21}(x)\sigma_2\otimes\sigma_1
    &\qquad \text{(IV) exotic interlayer interaction}
\end{cases}\label{classification}
\end{equation}
The coefficients $v_{\mu0}$ and $v_{\mu1}$ are uniquely fixed as functions of the matrix elements of $U$ and their derivatives.

In the current work, we will present a model of a coupler based on bilayer graphene. In this context, the potential terms (I) correspond to the electric field. Indeed, the electric field on the first layer would be  $v_{00}(x)+v_{30}(x)$, whereas on the second layer, it corresponds to $v_{00}(x)-v_{30}(x)$. The class (II) corresponds to the layer-symmetric ($v_{01}$) or layer-asymmetric ($v_{31}$) magnetic field perpendicular to layers  \cite{Katsnelson_book}. The third class (III) represents an AA-type interaction. The term $v_{10}$ is the classical AA-coupling. The strength of the coupling is given by a real-valued hopping parameter $t$. In the case of pristine AA bilayer graphene, its experimental value is $  t \approx 0.3-0.4~\mathrm{eV}$, see \cite{Rozhkov}.  The potential $v_{20}$ can be induced by a magnetic field parallel with the layers. The class (IV) corresponds to a rather exotic interlayer interaction that would exchange sublattices $A$ and $B$. All these interactions support perfect transmission in the normal direction. 
 
The presented construction of $V_{cpl}$ has one rather important by-product. The potential is reflectionless for Dirac fermions bouncing in the normal direction. Indeed, the system is unitarily equivalent to the free-particle system. Therefore, such system automatically possesses Klein tunneling. 

The construction of $V_{cpl}$ has an important and noteworthy consequence. The resulting potential is reflectionless for Dirac fermions incident normal to the interface. This follows from the fact that the system is unitarily equivalent to a free-particle model. As a result, the system necessarily exhibits Klein tunneling.

The free-particle Hamiltonian $H_{out}$ has a rather large family of symmetries that allow for different ways of polarization. They are formed by the matrices $e^{i\sigma_1\alpha}e^{i\sigma_3\beta}\otimes\sigma_{0}$ where $\alpha$ and $\beta$ are real parameters. Let us focus on three of them, denoted by $S_j$ whose explicit forms together with the corresponding basis of polarized states are as follows
\begin{equation}\label{eq:polarizace}
S_j=\sigma_j\otimes\sigma_0,\quad
\psi_{1,\pm}=\frac{1}{\sqrt{2}}\left(\Psi,\pm \Psi\right),\quad \psi_{2,\pm}=\frac{1}{\sqrt{2}}\left(\Psi,\pm i\,\Psi\right),\quad \psi_{3,\pm}=\frac{1}{2}\left(\Psi\pm\Psi,\Psi\mp\Psi\right).
\end{equation}
The polarized states satisfy
\begin{equation}
S_{j}\psi_{j,\pm}=\pm\psi_{j,\pm}.
\end{equation}
They form a basis in the corresponding representation space. In this basis, the coupler output $\psi_t$ can be written in general as 
\begin{equation}\label{outbases}
\psi_t=U(L)\psi=\sum_{\epsilon=\pm}c_{j,\epsilon}\psi_{j,\epsilon},
\end{equation}
where the polarization coefficients $c_{X,\pm}$, $X=\{1,2,3\}$, are complex constants in general. 

We aim for such a device that allows for a controllable change of the polarization coefficients. In this context, it is worth noting that the polarization with respect to $S_1$ is preserved by AA-interaction $v_{10}(x)\sigma_1\otimes\sigma_0$ whereas it gets changed by electric field $V_{cpl}=v_{30}(x)\sigma_3\otimes\sigma_0$. In AA-bilayer graphene, the eigenstates of $S_1$ correspond to the so called cone-polarized states as the polarized states belong to one of the two mutually shifted Dirac cones.
Polarization with respect to $S_3$ represents layer polarization. It is stable under the electric field $v_{30}(x)\sigma_3\otimes\sigma_0$ whereas it gets changed by the AA-interlayer interaction $v_{10}(x)\sigma_1\otimes\sigma_0$. Polarization associated with the symmetry operator $S_2$ is sensitive to both AA-interlayer interaction and electric field.

\section{AA-Bilayer coupler\label{section3}}
\begin{figure}[!h]
	\centering

    \hspace{1cm}
     \begin{subfigure}
{0.445\linewidth}
    \centering
    \includegraphics[width=\linewidth,height= 0.5\linewidth]{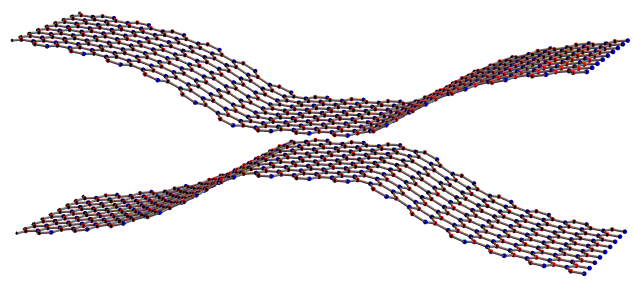}
    \caption{ Schematic view of coupler}
    \label{fig2d}
  \end{subfigure}
  \begin{subfigure}{0.48\linewidth}
    \centering
    \includegraphics[width=\textwidth,height =0.48\textwidth,keepaspectratio]{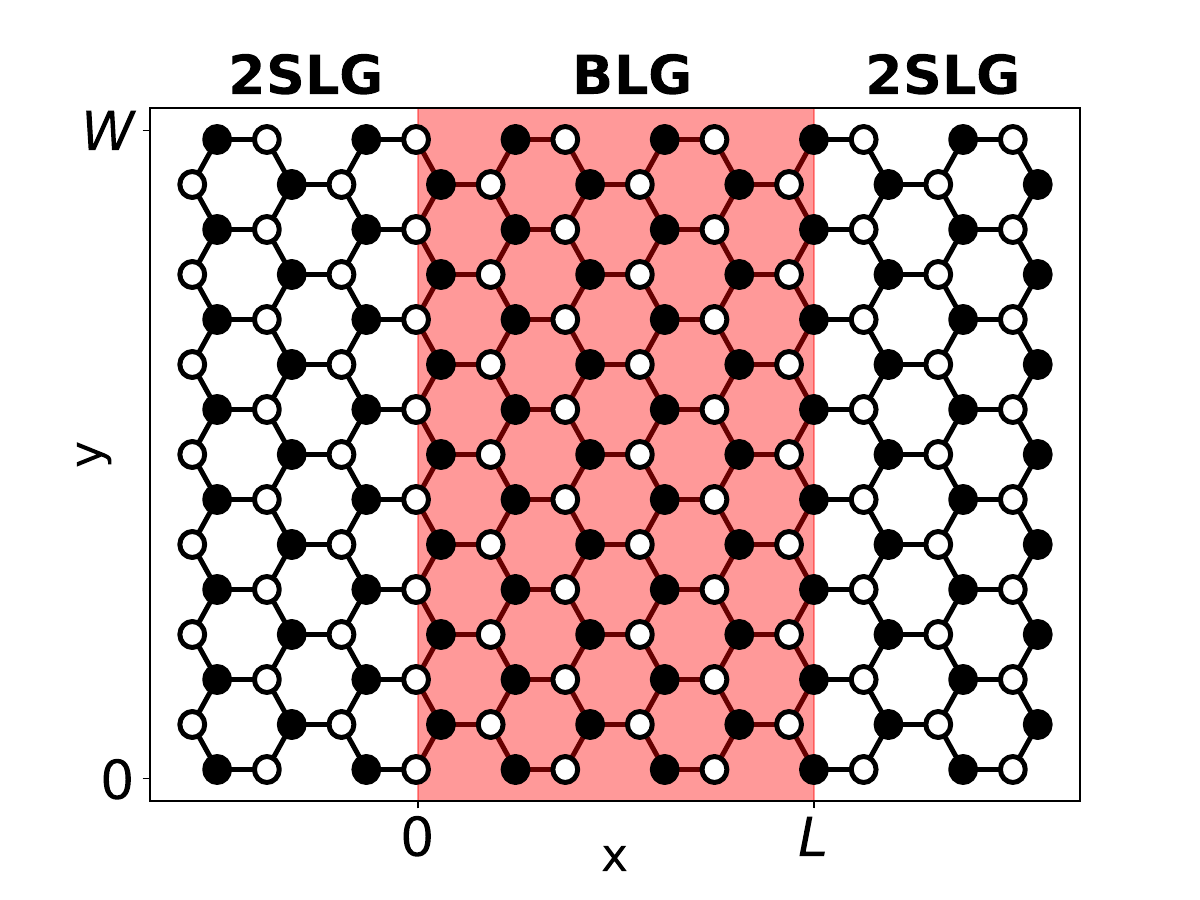}
    \caption{Coupler from top view}
    \label{fig2a}
  \end{subfigure}
	
	\caption{(\textit{Color online}) The model of  AA-stacked bilayer coupler. Two layers of graphene can get deformed such that they form the stripe of AA-stacking.}
	\label{fig:coupler}%
	\end{figure}

We shall discuss a model of a coupler based on parallel graphene nanorbibbons with armchair (AC) edges locally interacting via AA-interaction. The scheme of the system is presented in Fig.~\ref{fig:coupler}.  
Our goal is to analyze the situations where the coupler provides lossless to either upper or lower nanoribbon in a controllable manner. 

Dynamics of quasi-particles in the nanoribbons is described by two Dirac Hamiltonians that are locally coupled by interlayer interaction. Since the armchair boundary conditions mix the states from the two Dirac valleys, the total Hamiltonian has to take into account the valley degree of freedom.  It can be written in the following manner
\begin{equation}\label{eq:Ham_coupler}
\mathcal{H}\Psi=E\Psi,\quad \mathcal{H} =\sigma_0\otimes H= \sigma_0\otimes(H_{out}+V_{cpl}),
\end{equation}
where the first operator (identity matrix) in the tensor product acts on the valley index of the wave functions. 
The effective Hamiltonian $H$ acting in the single valley has two parts
\begin{equation}
 H_{out}=-i\sigma_0\otimes\sigma_1\partial_x-i\sigma_0\otimes\sigma_2\partial_y,\quad V_{cpl}=\sum_{j=1}^3 v_{j0}\, \sigma_j\otimes\sigma_0\,f(x).
\end{equation}
The potential consists of AA-interlayer interaction, electric field and magnetic field that is parallel with the coupler. The three couplings $v_{j0}$ are real constants and  $f(x)$ is the real function.   
The first Pauli matrix in the tensor product of $H$ acts on the layer-degree of freedom, whereas the second one acts on the pseudo-spin.
The order of the components of the wavefunction  is $\Psi=(\Psi_K,\Psi_{K'})$, where $\Psi_K = (\psi_{A1}, \psi_{B1},\psi_{A2},\psi_{B2})$ and  $\Psi_{K'} = (\psi_{B1}, \psi_{A1},\psi_{B2},\psi_{A2})$. Here, $\psi_{A1}$ is the sublattice $A$ and the layer 1 component, etc.  

The wave functions $\Psi(x,y)$ are required to be continuous  at $x=0,L$. In addition, they have to satisfy armchair boundary conditions at the edges of the nanoribbon. Explicitly, there should hold  
\begin{equation}
\Psi(0_-,y)=\Psi(0_+,y),\quad \Psi(L_-,y)=\Psi(L_+,y),\label{LBC}
\end{equation}
where $L$ is the length of the coupler and the subindex ${}_{\pm}$ denotes the limit from the left or from the right. The armchair edges are associated with the following boundary conditions
\begin{equation}\label{eq:BC}
    \Psi(x, 0) = M_1\Psi(x,0), \quad \Psi(x, W) = M_2\Psi(x, W),
\end{equation}
where $W$ is the width of the nanoribbons and  
\begin{equation}\label{BCM}
    M_1 = \sigma_2\otimes\sigma_0\otimes\sigma_1, \quad M_2 = \sigma_2\e{i\alpha\sigma_3}\otimes\sigma_0\otimes\sigma_1,
\end{equation}
see \cite{Brey_Fertrig}.  
Finite width of the nanoribbon implies quantization  of the momentum $k_y$. The parameter $\alpha\in\{0,\pm 2\pi/3\}$  is fixed by the width $W$ and  distinguishes the metallic and semiconducting AC nanoribbons \cite{Brey_Fertrig}. AC boundary conditions allow for an explicit solution of (\ref{eq:BC}) independently of the explicit form of the potential term. It is possible due to their "separability" that allows to construct the states that satisfy (\ref{eq:BC}) in the following manner (see Appendix \ref{sec:AC_projektor} for details),
\begin{equation}\label{eq:projected_states}
      \Psi(x, y) = (\mathbb{I} + M_1\mathcal{P}_y)e^{ik_yy}(\Psi_K(x), \Psi_{K'}(x)),
\end{equation}
where $\mathcal{P}_y\psi(x,y) = \psi(x, -y) $ and 
\begin{equation}\label{eq:AC_2pi/3_dispersion}
    k_y =\begin{cases} \frac{\pi m}{W},\quad\alpha=0,\\
    \frac{1}{W}\left(\pi m+\frac{\alpha}{2}\right),\quad \Psi_{K'}(x)\equiv 0,\\
    \frac{1}{W}\left(\pi m-\frac{\alpha}{2}\right),\quad \Psi_{K}(x)\equiv 0,\end{cases}  \quad  m\in \mathbb{Z}.
\end{equation}
It is worth highlighting that $e^{ik_y y}\Psi_{K(K')}(x)$ are arbitrary eigenstates of $H$ and explicit knowledge of $\Psi_{K(K')}(x)$ is not essential for making the boundary conditions (\ref{eq:BC}) satisfied. Therefore, we can limit ourself to calculation of the wave function $\Psi_K$ as $\Psi_{K'}$ can be obtained identically and the total wave function $\Psi$ is given by (\ref{eq:projected_states}).

\section{Narrow metallic coupler regime}\label{section4}
We assume that the armchair nanoribbons are in metallic configuration, i.e. $\alpha=0$ in (\ref{BCM}) and (\ref{eq:AC_2pi/3_dispersion}). Therefore, the subspace of $k_y=0$ is an admissible channel for dynamics. When the energy of the quasiparticles is such that $E<v_F\hbar\frac{\pi}{W}$, where we recovered the constants $\hbar$ and $v_F$, then the channel $k_y=0$ is the only admissible. We call this the narrow metallic coupler regime as the upper bound $v_F\hbar\frac{\pi}{W}$ is inverse-proportional to the width of the nanoribbons.  Considering  $E \sim 0.1~\mathrm{eV}$ as the energy scale of the Dirac regime, we get the following estimate for the $W$, 
\begin{equation}
    W < \frac{\pi\hbar v_F}{E} \sim 20~\mathrm{nm}.
\end{equation}

In this regime, the dynamics can be described by an effective Hamiltonian $\tilde{H}_{0}$ that can be mapped to free-particle dynamics by $U$, see (\ref{Vcpl}). 
We fix the unitary transformation $U$ as
\begin{equation}
      U(x) = \begin{cases}
         \mathbb{I},&x<0,\\
         \exp\left(-iQ\cdot \int_0^x f(x)dx\right)
         ,& x\in [0,L],\\
         \exp\left(-iQ\, \mathcal{L}
         \right) = \cos\left( q\, \mathcal{L}\right)-i\,\frac{Q}{q} \sin\left(q \,\mathcal{L}\right),& x>L,
     \end{cases}
 \end{equation}
where 
\begin{equation}\label{Qqomega}
Q=\sum_{j=1}^3 v_{j0}\sigma_j\otimes\sigma_1,\quad q=\sqrt{\sum_{j=1}^3v_{j0}^2},\quad \mbox{and}\quad \mathcal{L}= \int_{0}^Lf(x)dx,
\end{equation}
Then we get explicitly
\begin{eqnarray}
\tilde{H}_0&=&U(-i\sigma_0\otimes\sigma_1\partial_x)U^{-1}\nonumber\\
&=&-i\sigma_0\otimes\sigma_1\partial_x+f(x)\chi_{[0,L]}(x)\sum_{j=1}^3v_{j0}\,\sigma_j\otimes\sigma_0.\label{Uel}
\end{eqnarray}
where $\chi_{[0,L]}(x)=1$ for $x\in[0,L]$ and is zero otherwise.
The coupler is realized by electric $v_{30}$ and interlayer $v_{10}$ interactions. There is also parallel magnetic field $v_{20}$ that makes the hopping between the layers complex. In the calculation of (\ref{Uel}), it is important that $[\partial_xQ,Q]=0$, which is granted by the same functional dependence of the mentioned interactions. If the later commutator was non-vanishing, the unitary transform (\ref{Uel}) would also produce a potential from class (IV) in (\ref{classification}). 
There is Klein tunneling through the coupler as the absence of backscattering is inherited from the free particle system. 
It is worth mentioning that the vector potential of parallel magnetic field induces term $v_{20}$ from the original hopping $v_{10}$,  such that $v_{10} = t\cos(\tau)$ and $v_{20} = t \sin(\tau)$ where $\tau$ is the phase acquired by the particle when traveling between the layers. Therefore, there holds \footnote{In presence of parallel magnetic field, the interlayer hopping of the tight-binding Hamiltonian gets modified by the Peierls substitution 
\begin{equation}
     t \mapsto t_{12} = t\e{-i\frac{e}{\hbar}A_zd} \equiv t\e{-i\tau},   t_{21} = t_{12}^\star = t\e{i\tau}.
 \end{equation}
$t_{12}$ is hopping from layer one to layer two and $t_{21}$ is the reversed process. $d$ is the distance between layers.}
\begin{equation}
v_{10}^2 + v_{20}^2 = t^2.
\end{equation}

The eigenstates of $\tilde{H}_0$ are $\tilde{\psi}(x)=U(x)\psi(x)$ where $\psi(x)$ is an eigenstate of the free particle Hamiltonian. 
Consider an incoming state in the following general form
\begin{equation}\label{eq:incoming_state}
    \psi_{in}(x) \equiv 
    \frac{1}{\sqrt{2}}(\cos\gamma,e^{ia}\sin\gamma)\otimes(1,1)e^{ik_x x}, 
\quad \gamma \in \left[0,\frac{\pi}{2}\right],\quad a\in \mathbb{R}.
\end{equation}
The state transmitted through the coupler can be written as a linear combination in one of the basis of the polarized states,
\begin{equation}
\tilde{\psi}_t(x)=U(L)\psi(x)=c_{j+}\psi_{j+}(x)+c_{j-}\psi_{j-},\quad j=1,2,3
\end{equation}
where $\psi_{j\pm}$ and $c_{j\pm}$ are polarized states and the corresponding polarization coefficients, see (\ref{eq:polarizace}). It is straightforward to derive the explicit form of the coefficients $c_{j\pm}$. Nevertheless, one can deduce the type of polarization of the outgoing state by the ratio of $c_{3\pm}$. Indeed, taking into account (\ref{eq:polarizace}), 
\begin{equation}
\psi_t=\begin{cases}\psi_{1\pm},\quad \frac{c_{3-}}{c_{3+}}=\pm1,\\
\psi_{2\pm},\quad \frac{c_{3-}}{c_{3+}}=\pm i.\end{cases}
\end{equation}
The polarization coefficients $c_{3\pm}$ are explicitly
\begin{align}c_{3+}&=\left(\,\cos(q\mathcal{L}) - i\,\sin(q\mathcal{L})\, \frac{v_{30}}{q}\right)\cos(\gamma)-\frac{i\, e^{i(\alpha - \tau)}\, t\, \sin(q\mathcal{L})\, }{q}\sin(\gamma)
,\nonumber\\
c_{3-}&=\left(\,\cos(q\mathcal{L}) +i\,\sin(q\mathcal{L})\, \frac{v_{30}}{q}\right)\sin(\gamma)-\frac{i\, e^{i(\tau-\alpha)}\, t\, \, \sin(q\mathcal{L})}{q}\cos(\gamma)
.\label{TT} 
\end{align}

There are several important aspects that can be learned from these relations. The electric field acts against the effect of the coupler; indeed, $\psi_t\rightarrow\psi_{in}$ in the limit $v_{30}\rightarrow\infty$ (i.e. $q\rightarrow\infty,$ $v_{30}/q \to 1$). The change of polarization is mostly affected by the phase $q\mathcal{L}$. Complete change of $\psi_{3+}\rightarrow \psi_{3-}$, i.e. the layer flipping, occurs when $\cos q\mathcal{L}=0$. The magnetic field, represented by $\tau\neq0|_{mod\ 2\pi}$, causes periodic flipping of polarization between the states $\psi_{1\pm}$ and $\psi_{2\pm}$. Let us discuss these regimes in more detail below:
\begin{itemize}
\item \textbf{Layer-polarization converter} 
   
Let us assume that  for the case where the incoming state is polarized on the upper layer, $\psi_{in}=\psi_{3+}$, we have $\gamma=0$. Both magnetic and electric fields are switched off ($v_{30} = v_{20}=0$). The transport properties in this regime determined by the profile of the interlayer interaction $v_{10}\mathcal{L}$. There is a total transmission to the lower layer provided that $\cos q \mathcal{L} =0$, see Fig.\ref{couplerV}a). The corresponding "magic" values   are given by \footnote{This is related to magic length condition that appears in the literature \cite{Gonz_lez_2010}, \cite{Abdullah_2016}. Indeed, when $f(x)=f$ is a constant, then $\omega = q f L$. The relations $\cos q\mathcal{L}=0$ or $\sin q\mathcal{L}=0$  are then satisfied for "magic" lengths $L$.}
\begin{equation}\label{omegadown}
|v_{10}|
   = \frac{(2n+1)\pi}{2\mathcal{L}}\quad \Rightarrow \quad \psi_t=\psi_{3-},\quad n\in\mathbb{Z}.
\end{equation}
It can be a challenging task to fix either $|v_{10}|$ or $\mathcal{L}$ such that (\ref{omegadown}) is satisfied. It is worth mentioning in this context that the interlayer coupling can be tuned by changing the interlayer distance via applied pressure \cite{doi:10.1021/acs.nanolett.4c02035} up to $40\%$ of the original value of $v_{10}$.  This can lower the demand for precision in  preparation of (\ref{omegadown}).  


The electric field $v_{30}$ can be used to recover the original polarization, i.e. $\psi_t=\psi_{3+}$. It follows directly from (\ref{TT}). Indeed,  when switching on and increasing $v_{30}$, $|c_{3+}|$ converges to one whereas $|c_{3-}|$ goes to zero. There are resonant values of the electric field $v_{30}$ where $|c_{3+}|=1$. They are specified implicitly by equation 
\begin{equation}\label{eq:switch}
\sin\sqrt{v_{10}^2+v_{30}^2}\,\mathcal{L}=0,
\end{equation}
Therefore, the electric field $v_{30}$ can be used to cancel the effect of the coupler, see Fig.\ref{couplerV}a). 

It is worth mentioning that the choice (\ref{omegadown}) is not ideal for reaching perfect cone-polarization $|c_{1\pm}|=1$,  the maximal values of the $|c_{1\pm}|$ vary both with $v_{30}$ and $n$. Still, we can get reasonable approximation of these values dependently on the explicit choice of $n$ in (\ref{omegadown}) and $v_{30}$ see Fig.\ref{couplerV}b) where $|c_{1\pm}|>0.95$ for $n=1$. The case of perfect conversion $c_{3+}\rightarrow c_{1\pm}$ will be discussed in the next paragraph.

\begin{figure}[H]
 	\centering\includegraphics[width=.4\linewidth,
    trim=0cm 1.cm -0.1cm 1cm,clip]{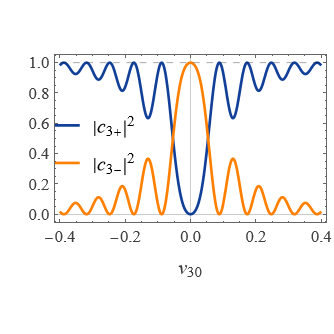} \includegraphics[width=.4\linewidth,
    trim=0cm 1.cm -0.1cm 1cm,clip]{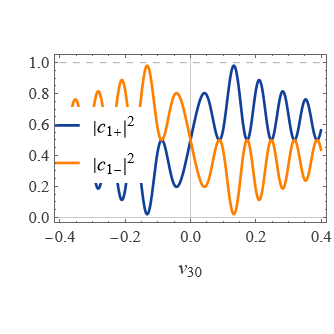}
	\caption{Polarization coefficients $|c_{3\pm}|^2$ (left) and $|c_{1\pm}|^2$ (right) in dependence on the electric  field $v_{30}$. We fixed $|v_{10}|=0.1$, $n=1$ that corresponds to $\mathcal{L}\sim47.12$.}%
	\label{couplerV}%
\end{figure}

\item \textbf{Layer-cone-polarization converter}:\\

The coupler can be set such that it converts layer polarized states into cone-polarized ones. Let us have layer polarized state
\begin{equation}
\psi_{in}=\frac{1}{\sqrt{2}}(1,0)\otimes(1,1)\,e^{ikx}
\end{equation}
We fix 
\begin{equation}t=\frac{(2n+1)\pi \cos\gamma}{2\mathcal{L}},\quad v_{30}=\frac{(2n+1)\pi \sin\gamma}{2\mathcal{L}}.
\end{equation}
Then there holds
\begin{equation}
\psi_t=U(L)\psi_{in}\sim(\sin\gamma,-e^{i\tau}\cos\gamma)\otimes(1,1)\,e^{ikx}.
\end{equation}
Fixing $v_{30}=t$ corresponds to $\gamma=\pi/4$. Then the polarization of the outgoing state can be tuned to any of $\psi_{1\pm}$ and $\psi_{2\pm}$ by parallel magnetic field,
\begin{equation}
\psi_t=\begin{cases}\psi_{1-},\quad\tau=0,\\\psi_{1+},\quad \tau=\pi,\\\psi_{2-},\quad\tau=\pi/2,\\\psi_{2+},\quad\tau=-\pi/2,\end{cases}
\end{equation}
see Fig.\ref{coupler2} for illustration. It is worth mentioning that conversion of polarization of light on metasurfaces was discussed in  \cite{PhysRevApplied.22.054028}.

\begin{figure}[H]
 	\centering
    \includegraphics[width=.4\linewidth,
    trim=0cm 1.cm -0.1cm 1cm,clip]{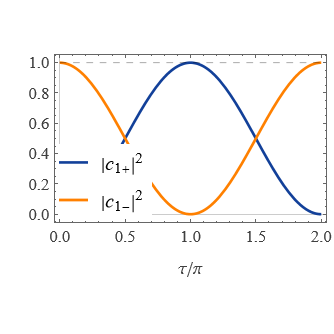} 
\includegraphics[width=.4\linewidth,
    trim=0cm 1.cm 0cm 1cm,clip]{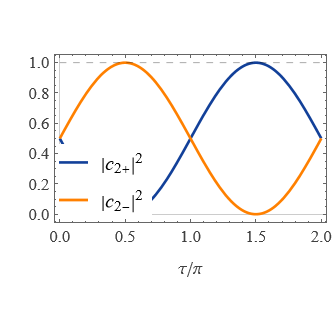} 
	\caption{Oscillation of polarization coefficients $|c_{1\pm}|^2$ and $|c_{2\pm}|^2$ as function of parallel magnetic field. We fixed $|v_{10}|=0.1$.} 
	\label{coupler2}%
\end{figure}

\item{\textbf{Interferometer (inverted cone-layer polarizer)\\}}
Let us assume that explicit polarization of the incoming state is unknown. It has the following form
\begin{equation}
\psi_{in}=\frac{1}{\sqrt{2}}(\cos\gamma,e^{i\beta}\sin\gamma)\otimes(1,1)e^{ikx}
\end{equation}
where $\gamma$ and $\beta$ are are unspecified parameters. The parameters can be determined by tuning $t$ and $v_{30}$ such that the outgoing state is localized on upper layer only. Indeed, fixing
\begin{equation}
v_{30} =\frac{\pi (1 + 2n)\cos\gamma}{2\mathcal{L}}, \quad
t=\frac{\pi (1 + 2n)\sin\gamma}{2\mathcal{L}}, \quad
\tau=\beta.\end{equation}
there holds $|c_{3+}|=1$ and $\tan\gamma=\frac{t}{v_{30}}$. This is actually the same device as in the previous paragraph, yet inverted.

\end{itemize}


\section{Wider couplers\label{section5}}
Let us focus on the situation where the narrow coupler regime is no longer justified as there are admissible additional channels with $k_y\neq0$ for transmission. This can happen due to larger energies of incoming particles or wider width of the coupler that makes quantization of $k_y$ finer. 

\begin{figure}[h]
  \centering

    \includegraphics[width=.35\linewidth]{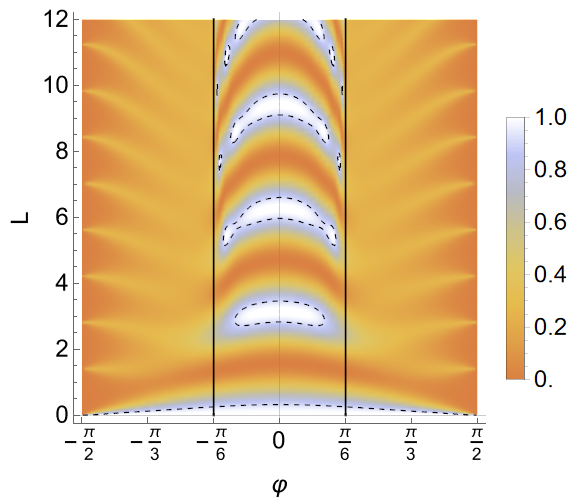}
    \includegraphics[width=.35\linewidth]{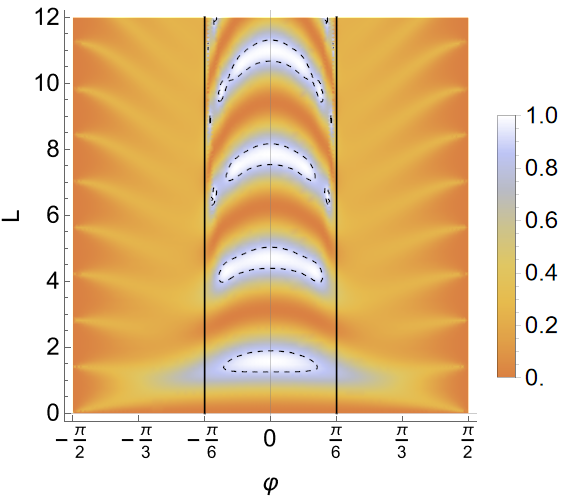}
    
  
 

  \caption{Transmission on the layer one $\abs{c_{3+}}^2$  (left) and the layer two $\abs{c_{3-}}^2$ (right) as a function of $L$ and  $\varphi$. The area between dashed curves has transmission greater than $95\%$. Black vertical lines symbolize the critical angle $\varphi_c^{-}$. }
\label{fig2}
\end{figure}

\begin{figure}[H]
 	\centering
 	\subfloat[\centering $|c_{3+}(\varphi, E)|^2, v_{30} = 0$\label{fig3a}]{{\includegraphics[height =7cm,width = 6cm,keepaspectratio]{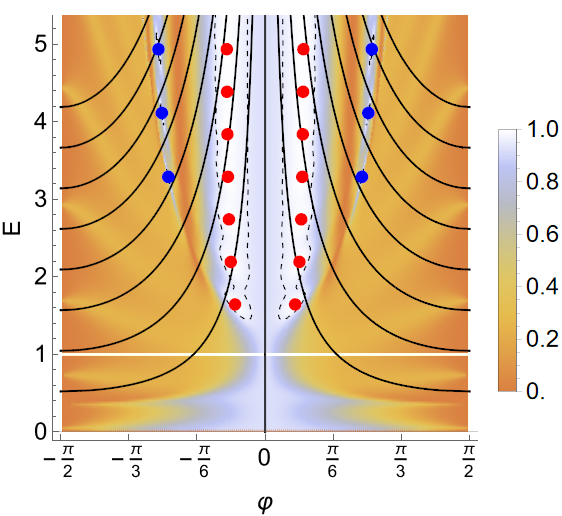} }}%
 	\subfloat[\centering $|c_{3-}(\varphi, E)|^2, v_{30}=0$ \label{fig3b}]{{\includegraphics[height =7cm,width = 6cm,keepaspectratio]{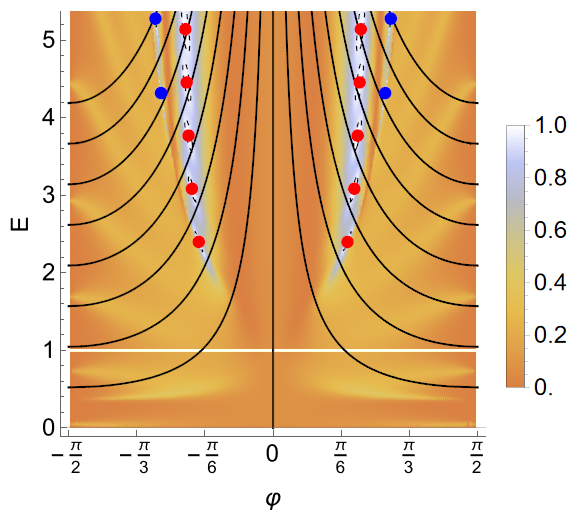} }}
    \\
    \subfloat[\centering $|c_{3+}(\varphi, E)|^2, v_{30}\neq 0$\label{fig3c}]{{\includegraphics[height =8cm,width = 6cm,keepaspectratio]{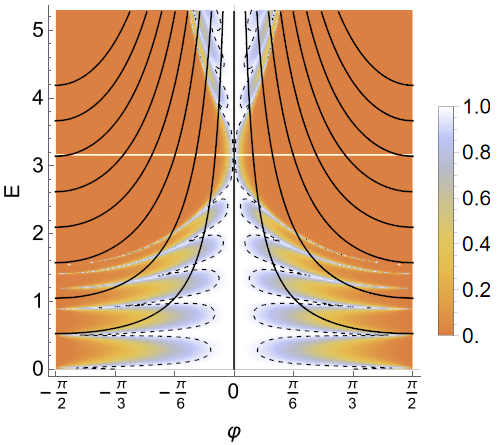} }}%
 	\subfloat[\centering $|c_{3-}(\varphi, E)|^2, v_{30} \neq 0$ \label{fig3d}]{{\includegraphics[height =8cm,width = 6cm,keepaspectratio]{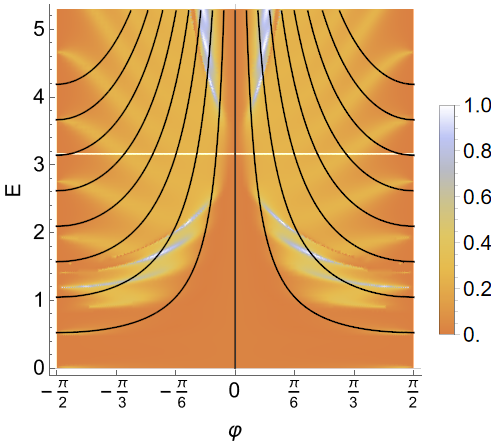} }}%
 	\caption{Transmission on the layer one $|c_{3+}|^2$ and the layer two $|c_{3-}|^2$ as a function of the energy $E$ with $v_{30}=0$ (upper line) and with $v_{30}\neq 0$ (lower line). We set $W = L=6, t=1, v_{30}=3$. The white line indicates  $E = \sqrt{v_{30}^2 + t^2}$. The white area between dashed curves has transmission greater than $95\%$. Red points have $l=2$, blue points have $l=3$, see (\ref{eq:cosT1}) and (\ref{eq:cosT2}). The solid black lines are incidence angles $\varphi_m$ given by \eqref{eq:quant_uhly} in the metallic armchair terminated coupler, $m\in \{0,1,2\dots 8\}$. 
    }
 	\label{fig3}%
 \end{figure}

Let us briefly discuss the scattering properties of the coupler, where  
\begin{equation}v_{10}= t = const.,\quad  f(x) = 1\Longrightarrow \mathcal{L}=L,\label{constants}\end{equation}
with focus on the layer polarization.
We make the following ansatz for the scattering state with incoming wave localized on the first layer,
\begin{equation}
\Psi=\begin{cases}\begin{pmatrix}
 		\psi_{E,k_x}\\
 		0
 	\end{pmatrix}+r_{3+}\begin{pmatrix}
 		\psi_{E,-k_x}\\
 		0
 	\end{pmatrix}+r_{3-}\begin{pmatrix}
 		0\\
 		\psi_{E,-k_x}
 	\end{pmatrix}=\Psi_{in}+r_{3+}\Psi_{r}^{(+)}+r_{3-}\Psi_{r}^{(-)},& x<0,\\
    \sum_{\epsilon=\pm}\left(A_\epsilon\begin{pmatrix}
 		-\epsilon\psi_{E+\epsilon t,q_\epsilon}\\
 		\psi_{E+\epsilon t,q_\epsilon}
 	\end{pmatrix}+B_\epsilon\begin{pmatrix}
 		-\epsilon\psi_{E+\epsilon t,-q_\epsilon}\\
 		\psi_{E+\epsilon t,-q_\epsilon}
 	\end{pmatrix}\right),& x\in[0,L],
\\
    c_{3+}\begin{pmatrix}
 		\psi_{E,kx}\\
 		0
 	\end{pmatrix}+c_{3-}
\begin{pmatrix} 		0\\
 		\psi_{E,kx}
 	\end{pmatrix}=c_{3+}\Psi_{3+}+c_{3-}\Psi_{3-},& x>L.
\end{cases}\label{ansatzpsi}
\end{equation}
Here $q_\pm$ is implicitly defined as $q_{\pm}^2+k_y^2=(E\pm t)^2$. The spinors $\psi_{E,k_x}$ read explicitly as follows
\begin{equation}\label{eq:psi_in}
 	 \psi_{E,k_x}=e^{i(k_xx +k_yy )}\begin{pmatrix}1 \\
 		\xi e^{i\varphi}\end{pmatrix},\quad \mathrm{tan}(\varphi) = \frac{k_y}{k_x},\quad E^2 = k_x^2 + k_y^2,\quad \xi = \sgn{E}. 
 \end{equation}
Here, $k_y$ is fixed as in (\ref{eq:AC_2pi/3_dispersion}). 
Substituting into  (\ref{LBC}), we get a system of eight linear equations for unknown coefficients $r_{3\pm}, c_{3\pm}$, $A_{\pm}$ and $B_{\pm}$ that must hold for any real $y$. Instead of presenting the explicit analytical solution of the equations (\ref{LBC}) given in terms of rather extensive formulas, we provide the explicit values of the transmission amplitudes $c_{3\pm}$ as function of $L$ and $\phi$ in Fig.~\ref{fig2} and $E$ and $\phi$ in Fig.~\ref{fig3}.

Conservation of $k_y$ outside and within the coupler implies that the particles with higher energy can pass through the coupler at a narrow angles of incidence only. Indeed, there holds
 \begin{equation}\label{eq:zachovani_ky}
 	 \abs{E}\sin(\varphi) = \abs{E\pm t}\sin(\theta^{\pm}),\quad \tan\theta^{\pm}=\frac{k_y}{q_{\pm}}.
 \end{equation}
 In particular, when $\varphi$ is such that there holds 
 \begin{equation}\label{eq:damping}
 	\abs{\sin(\theta^{\pm})} = \abs{\frac{{E}}{{E\pm t}}\sin(\varphi)}<1,
 \end{equation}
then we get real-valued $q_{\pm}$ and oscillating solutions in the coupler. The transmitted wave is exponentially damped otherwise, see Fig.~\ref{fig2}. 
There is no damping for $\abs{E}<\frac{t}{2}$ as \eqref{eq:damping}  is satisfied for any $\varphi\in(-\pi/2,\pi/2)$.  For $\abs{E}>t/2$, the wave function is exponentially damped at incoming angles greater than $\varphi_c^{\pm}$, where $\sin(\varphi_c^{\pm}) = \frac{\abs{E\pm t}}{\abs{E}}$.
When $E=t/2$, $\varphi_c^{\pm}\rightarrow 0$ for $E=t/2$ and the transmission forms a "bottle-neck", see  in the Fig.\ref{fig3a}, Fig.\ref{fig3b}. 
Taking into account (\ref{constants}), the magic phase (\ref{omegadown}) can be rewritten into the condition for the magic lengths $L_{\uparrow} = \frac{n}{t}\pi, L_{\downarrow} = \frac{2n+1}{t}\pi$. For $\phi=0$, the magic lengths are associated with perfect transmission to either layer 1 or layer 2, see Fig.~\ref{fig2} for $\phi=0$.

For $k_y\neq 0$, perfect transmission occurs for discrete configurations of energies and angles, corresponding to Fabry-Pérot resonances. 
First, let us specify the Fabry-Pérot resonances in the first layer. We fix $r_{3\pm}=0$ and $c_{3-}=0$ in the ansatz  (\ref{ansatzpsi}). The matching conditions can be satisfied provided that the following two equalities are satisfied simultaneously, 
\begin{equation}\label{eq:widthAA1}
   L\sqrt{(E- t)^2 - E^2\sin^2\varphi}=\pi n\quad\mbox{and}\quad L\sqrt{(E+ t)^2 - E^2\sin^2\varphi} = \pi (n+2l),
\end{equation}
where $n$ and $l$ are integers. Assuming that the width $L$ is a fixed parameter of the system, we can solve the equations above with the following energy $E_{\uparrow}(n,l)$ and angle $\varphi_{\uparrow}(n,l)$ of the perfect transmission to the first layer, $|c_{3+}(E_{\uparrow},\varphi_{\uparrow})|=1$,
\begin{equation}\label{eq:cosT1}
    E_{\uparrow}(n,l)= \left(\frac{\pi}{L\sqrt{t}}\right)^2(n+l)l, \quad \cos^2(\varphi_{\uparrow}(n,l)) = \frac{L^2t^2}{\pi^2}\left(\frac{n^2}{l^2(n+l)^2}+\frac{2}{(l+n)^2}+\frac{2\,n\,l-\frac{L^2t^2}{\pi^2}}{l^2(l+n)^2}\right).
\end{equation}
We can get the energies and angles of perfect transmission $|c_{3-}(E_{\downarrow},\phi_{\downarrow})|=1$ to the second layer similarly. They are 
\begin{align}\label{eq:cosT2}
E_{\downarrow}(n,l)&=E_{\uparrow}\left(n,l+\frac{1}{2}\right),\quad \cos^2(\phi_{\downarrow}(n,l))=\cos^2(\phi_{\uparrow}(n,l+1/2)).
\end{align}
The pairs $(\varphi_{\uparrow}, E_{\uparrow})$ or $(\varphi_{\downarrow}, E_{\downarrow})$ define the red and blue points in Fig.~\ref{fig3a}, \ref{fig3b}  where the transmission is 100\%.
It follows from (\ref{eq:cosT1}) and (\ref{eq:cosT2}) that 
the angles of these perfect resonant transmissions converge to a constant value in the limit of large energies. Indeed, taking the limit, i.e. $n\gg 1$, we get 
\begin{equation}\label{eq:AA_tr_angle_highE}
     \lim_{n\rightarrow \pm\infty}\cos{\varphi_{\uparrow}} = \left|\frac{tL}{\pi l}\right|, \qquad \lim_{n\rightarrow \pm\infty}\cos{\varphi_{\downarrow}} = \left|\frac{tL}{\pi \left(l+\frac{1}{2}\right)}\right|.
 \end{equation}
These formulas explain that points of perfect transmission form branches in $|c_{3\pm }(\varphi, E)|^2$ labeled by $l$, see Fig.~\ref{fig3a}, \ref{fig3b}.  

The transmission angles are functions of the transversal momentum $k_m$ that is quantized. The admissible values for given $E$ and $m$ are
\begin{equation}\label{eq:quant_uhly}
    \tan{\varphi_m} = \frac{k_y}{k_x} = \frac{\frac{1}{W}\left(\pi m\pm\frac{\alpha}{2}\right)}{\sqrt{E^2 - \left(\frac{1}{W}\left(\pi m\pm\frac{\alpha}{2}\right)\right)^2}},\quad m\in \mathbb{Z}.
\end{equation}
The admissible angles $\varphi_m$ 
are represented by solid black curves in Fig.~\ref{fig3}. The continuous transmission profile $|c_{3\pm}(\varphi, E)|^2$ is reduced to the allowed set of values $|c_{3\pm}(\varphi_m(E), E)|^2$. The quantized angles $\varphi_m$ can coincide with the perfect transmission angles $\varphi_{\uparrow}$ or $\varphi_{\downarrow}$, see Fig.~\ref{fig3a} and Fig.~\ref{fig3b}. This can have major impact on the transmission to layer 2 that is very low otherwise. 
Thus, a properly tuned armchair-terminated coupler can selectively transport Dirac fermions with specific angle of incidence and energy from layer 1 to  layer 2.

At the end of the section, let us specify the typical length of the coupler. Let us denote the real hopping $t = v_{10}  \approx 0.35~eV$. The product $v_{10}L = tL$ should be $tL \sim p\pi, p \in \mathbb{N} $. Restoring the units we have $p\pi\sim tL/(\hbar v_F) \approx L\frac{0.35}{0.66~\mathrm{nm}}$. The lengths should be chosen in a nanometer scale 
\begin{equation}
    L \sim 6p~\mathrm{nm}, p\in \mathbb{N}.
\end{equation}

\section{Conductivity in low-energy regime: Landauer approach 
\label{section6}}
The conductance $G_{+}$ on layer 1 and $G_{-}$ on layer 2 is determined by the transmission probability of Dirac fermions through the scattering region. The conductance can be calculated via Landauer formula\footnote{The polarization and transmission coefficients coincide. }
\begin{equation}
    G_{\pm}
    =
       G_0\sum_{k_y}^{\mathrm{open}}
    |c_{3\pm}(E_F,k_y)|^2,\quad G_0= \frac{e^2 g_s g_v}{2\pi\hbar}.
\end{equation}
The transverse momentum $k_y$ is quantized \eqref{eq:AC_2pi/3_dispersion}. The range of indices $m$ that correspond to open transport channels is determined by 
\begin{equation}\label{eq:kyAC_omezeni}
    \abs{k_y} = \abs{\frac{1}{W}\left(m\pi \pm \frac{\alpha}{2}\right)} \leq \frac{E_F}{\hbar v_F}. 
\end{equation}
In a semiconducting armchair nanoribbon, there is no conductace when $E_F$ falls into a gap as there are no open channels there. 

In the regime of narrow metallic coupler, the channel $m=0$ is the only available for low energies with $E_F< \pi\hbar v_F/W$. 
There is perfect transmission through the coupler, and, consequently,  $G=G_++G_-=G_0$. Actually, the conductivity $G_{\pm}$ is proportional to $|c_{3\pm}|^2$ were $c_{3\pm}$ is given in (\ref{TT}). Discussion related to (\ref{eq:switch}) applies  here directly. Therefore, we can conclude that 
tuning the perpendicular electric field enables the switching of Dirac fermions between the two layers, and dampen $G_-$ effectively for larger values of $v_{30}$, see Fig.\ref{couplerV}.

\begin{figure}[H]
 	\centering
 	\subfloat[\centering semiconducting $\alpha=2\pi/3$, narrow, $W = 1$\label{fig6a}]
{{\includegraphics[height =5.5cm,width = 5.5cm,keepaspectratio]{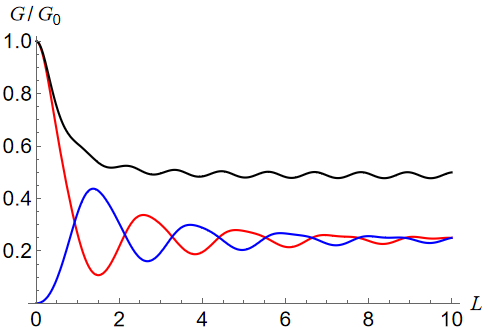} }}
 	\subfloat[\centering Semiconducting AC type $\alpha = 2\pi/3$ \label{fig:switchC}]{{\includegraphics[height =5.5cm,width = 5.5cm,keepaspectratio]{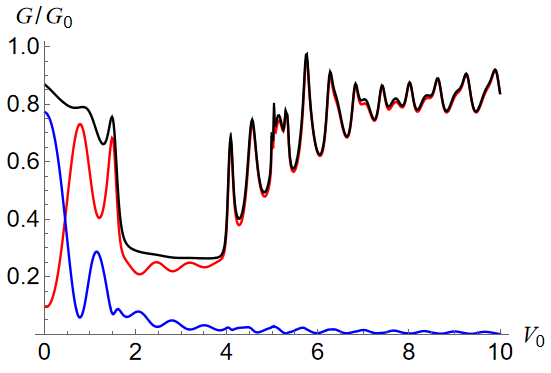} }}%
 	\caption{The conductance of AA bilayer coupler with semiconducting AC edges . The parameter are $v_{10} = t=1, W=1.03, E_F=3 $ and magic length $L = L_{\downarrow} = 3/2\pi$. The electric field is here denoted $v_{30} \equiv V_0$. }%
 	\label{fig6b}%
 \end{figure}

In contrast, a semiconducting-type coupler does not support the $k_y = 0$ channel. The modes $k_y\neq 0$ exhibit low transmission. This leads to a pronounced suppression of conductivity in the semiconducting-type coupler which cannot reach $G_0$ for $L>0$ (see Fig.~\ref{fig6b}a). On this figure, the semiconducting AC coupler cannot provide perfect switching mechanism. 
For an electric field such that $\sqrt{v_{30}^2 + t^2} \approx E_F$, the transmission is restricted to a "bottle-neck", a small angular range $\varphi \in (-\delta, \delta)$, see Fig.~\ref{fig3c}. Here, it reaches the maximal value $|c_{3+}|^2 + |c_{3-}|^2 = 1$. However, semiconducting coupler supports $\varphi \notin (-\delta, \delta)$ and the total transmission is strongly suppressed, see Fig. \ref{fig6b}b.  For high electric potential, the layer 2 conductance is suppressed as predicted in Fig.~\ref{couplerV}a.



\section{Conclusion}

In this article, we investigated the design of a quantum coupler that enables deliberate manipulation of the polarization of quantum states. Polarization is associated with a discrete internal degree of freedom. We provided a general framework based on symmetry consideration for controlling it within a quantum coupler. In section~\ref{section2}, we focused on regimes in which Klein tunneling ensures lossless transmission of quantum states. We demonstrated that the dynamics of such devices is effectively governed by free‑particle motion. In this sense, the present work generalizes the approach of \cite{Jakubsky2011} where Klein tunneling was interpreted as free‑particle dynamics in disguise. 

To illustrate the general results, an explicit model of the coupler was considered. It was formed by two graphene armchair nanoribbons brought locally close enough to allow  AA-interlayer interaction.  

In section~\ref{section4}, we concentrated on the narrow metallic‑coupler regime, where a single channel of particles incident perpendicularly on the coupler is relevant. We analyzed transmission into various polarized states as a function of the interlayer interaction, as well as external electric and magnetic fields where the later one was parallel with the nanoribbon (\ref{TT}). Our results show that tuning the interlayer coupling enables precise control over the transmitted polarization, while the electric field tends to suppress the coupler’s effect. Several operational regimes were identified, including a layer‑polarization converter, a layer–cone‑polarization converter, and an interferometric configuration capable of probing the polarization properties of the incoming state.

Beyond the narrow‑coupler regime, we also explored transport at higher energies in section~\ref{section5}. We determined the critical angles of incidence associated with a rapid drop in transmission once exceeded. Furthermore, we analytically identified the combinations of energies and incidence angles that give rise to Fabry–P\'erot resonances (\ref{eq:cosT1}), (\ref{eq:cosT2}), enabling perfect transmission even for oblique incidence. Conductance in narrow-metallic coupler regime as well as in the case of semiconducting armchair nanoribbons was computed in the section~\ref{section6}.

Although the article primarily focuses on polarization with respect to the layer degree of freedom, the conceptual framework introduced in section~\ref{section2} namely, the interpretation of Klein tunneling as free‑particle dynamics,  is broadly applicable. It can be extended to systems where polarization with respect to other internal degrees of freedom is relevant, such as spin. The possibility of using this approach for qubit manipulation represents an intriguing direction for future research, but it lies beyond the scope of the present work.

\section*{Acknowledgement}
VJ acknowledges the assistance provided by the Advanced Multiscale Materials for Key Enabling Technologies project, supported by the Ministry of Education, Youth, and Sports of the Czech Republic. Project No. CZ.02.01.01/00/22 008/0004558, Co-funded by the European Union. AMULET project. PČ was supported by   the Grant Agency of the Czech Technical University in Prague, grant no. SGS25/165/OHK4/3T/14.

\appendix
 \section{Quantization of momentum: armchair edges}\label{sec:AC_projektor}
Given the Hamiltonian
\begin{equation}
\mathcal{H} = \sigma_0 \otimes H ,
\label{eq:H_app_PRB}
\end{equation}
where $\sigma_0$ acts in the valley space and $H$ is in Eq.~\eqref{eq:Ham_coupler} and the component ordering is written as $\Psi = (\Psi_K,\Psi_{K'})$. 
The armchair boundary conditions are given by \eqref{eq:BC}.
We employ the projector construction~\cite{VJ_armchair}. Since
\begin{equation}
[\mathcal{H},M_1\mathcal{P}_y]=0,
\end{equation}
with $\mathcal{P}_y\psi(x,y)=\psi(x,-y)$, an eigenstate satisfying the boundary condition at $y=0$ is obtained as
\begin{equation}
\psi(x,y)=(\mathbb{I}+M_1\mathcal{P}_y)\Phi(x,y),
\label{eq:proj_PRB}
\end{equation}
where $\Phi$ is an eigenstate of $\mathcal{H}$. The second boundary condition reduces to
\begin{equation}\label{eq:druha_podm_AC}
\Phi(x,-W)=M_1^{-1}M_2\Phi(x,W).
\end{equation}
Assuming separability,
\begin{equation}
\Phi(x,y)=e^{ik_y y}
\begin{pmatrix}
\phi_K(x)\\
\phi_{K'}(x)
\end{pmatrix},
\end{equation}
we obtain from \eqref{eq:druha_podm_AC}
\begin{equation}
\begin{pmatrix}
\phi_K(x)\\
\phi_{K'}(x)
\end{pmatrix}
=
e^{2ik_yW}
\begin{pmatrix}
e^{i\alpha}\sigma_0\otimes\sigma_0 & 0\\
0 & e^{-i\alpha}\sigma_0\otimes\sigma_0
\end{pmatrix}
\begin{pmatrix}
\phi_K(x)\\
\phi_{K'}(x)
\end{pmatrix}.
\end{equation}
This yields the quantization condition
\begin{equation}
\begin{aligned}
&\alpha\neq0:\quad 
k_y=\frac{1}{W}\left(m\pi+\frac{\alpha}{2}\right), \quad \phi_{K'}(x)\equiv0, \\
&\hspace{1.4cm}\quad  
k_y=\frac{1}{W}\left(m\pi-\frac{\alpha}{2}\right), \quad \phi_K(x)\equiv0, \\
&\alpha=0:\quad 
k_y=\frac{m\pi}{W},
\end{aligned}
\qquad m\in\mathbb{Z}.
\label{eq:ky_quant_PRB}
\end{equation}
The conditions $\phi_K\equiv0$ or $\phi_{K'}\equiv0$ do not imply valley polarization. The full wavefunction,
\begin{equation}
\psi(x,y)=(\mathbb{I}+M_1\mathcal{P}_y)e^{ik_y y}\phi(x),
\end{equation}
contains components from both valleys due to the explicit boundary-induced valley mixing provided by matrix $M_1$. The states of metallic case have valley degeneracy two $g_v = 2$, in contrast to valley non-degenerate $g_v=1$ semiconducting case.

\bibliography{ref.bib}
\bibliographystyle{unsrt}

\end{document}